# SaaS CloudQual: A Quality Model for Evaluating Software as a Service on the Cloud Computing Environment

Dhanamma Jagli, Seema Purohit and N. Subash Chandra

**Abstract** The cloud computing is a key computing approach adopted by many organizations in order to share resources. It provides Everything As-A-Service (XaaS). Software-As-A-Service is an important resource on the cloud computing environment. Without installing any software locally, service user can use software as a utility. And enjoy the benefits of SaaS model. Hence SaaS usage is increased drastically, the demand for selecting quality is also increased. This paper presents a novel quality model intended for evaluating software as a service (SaaS), depending on the key features of Software as a service. Because SaaS key features are playing critical role in the quality and differentiating from conventional software quality.

**Keywords** Cloud computing · Software-As-A-Service (SAAS) · Service quality · Software quality

## 1 Introduction

The cloud computing as a sophistication in computing epitomizes vigorously scalable and regularly virtualized resources, which are delivered as a service through a web browser via Internet. In the cloud computing environment, everything such as network, infrastructure, platform, software or application is available as a service. The unique type of cloud service is Software-As-A-Service (SaaS).

D. Jagli (✉) · S. Purohit · N. Subash Chandra
JNTU Hyderabad, Hyderabad, India
e-mail: dsjagli.vesit@gmail.com

S. Purohit
e-mail: supurohit@gmail.com

N. Subash Chandra
e-mail: subhashchandra_n@yahoo.co.in

S. Purohit
Kirti College, University of Mumbai, Mumbai, India

It stands commonly used service by many customers and service providers to provide benefits to service their customers. In order to understand these benefits, evaluating the quality of SaaS on cloud has become extremely essential due the increased demand. This quality model further helps to manage quality at the top level as per the evaluation results. Existing conventional quality models are not competent enough for providing all Software service-specific features [1]. In this paper, initially, the description of conventional quality model for software and service is given separately. Further the paper describes about the proposed model followed by results and discussion. Finally, it concludes with future scope.

## 2 Proposed Quality Model

In this paper an innovative quality model is suggested to assess quality of software as a service on cloud, produced around quality attributes.

(a) Software Quality Model-ISO/IEC 9126 (25010): (PERFUM)

The ISO/IEC standard quality models are of two types: product quality model and quality in use model. Based on the static and dynamic properties a product quality model is further subdivided as internal product quality and external product quality model. Quality in use model is a system model used to relate the usage of a product to the context of the use. The ISO/IEC Standard quality model later modified as 25010 quality model to know software products quality. This model has identical internal and external quality characteristics and sub characteristics. The variance is in the quality measures. Quality in use has no sub characteristics [2]. According to ISO/IEC 9126 standards, software quality can be assessed using six characteristics. The software product quality model has six characteristics and 24 sub characteristics. In this paper, proposed quality model evaluates software product, the metrics or measures of the software product quality similar to ISO/IEC 9126 [3]. The standard attributes have been shown in the Fig. 1.

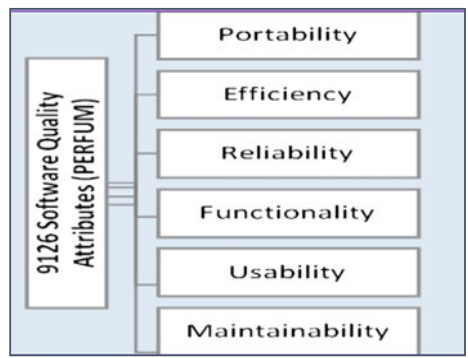

**Fig. 1** ISO/IEC9126 quality model

Due to the gap between conventional and clouding computing paradigms, the traditional quality models, based on ISO 9126 standards, are inadequate for assessing quality of SaaS [1]. They do not support efficient and effective evaluation of cloud computing explicit quality features. A quality model which can completely evaluate the needs of SaaS on cloud yet to arise [1]. Hence, rigorous efforts are being made at developing a quality model needed for measuring Software-As-A-Service on the cloud computing environment.

(b) Service Quality Model: (RATER)

In 1988, Parasuraman and his team introduced a standard quality model for service evaluation with five dimensions were used in order to evaluate the service quality provided to any service users. That quality method is called as SERVEQUAL, which has become a very popular model for service quality. "The service qualities emphasized are reliable, assurance, responsiveness, empathy and tangible" [4] as shown in the Fig. 2. Service user satisfaction is an increasing worry of any businesses all through the world [5]. In order to Evaluate SaaS quality, it also required to evaluate service quality.

(c) Key Features of Software-As-A-Service(SaaS)

The quality of SaaS includes quality software product plus quality of service. The key features of SaaS are critical and play an important role while describing quality of software as a service [6]. Identified Seven key features of SaaS are identified as shown in the Fig. 3.

- *Multi-tenant*: Multi-tenant means acceptable to the proposal which has profitable clarifications to multiple end user. The Software as a service essentially attains multi-tenancy. That is it has a capability to fulfil multiple end-users in parallel built on a solitary application instance. The service users want same functionality. Generally, multi-dimensional QoS parameters are response time, throughput and availability [7].

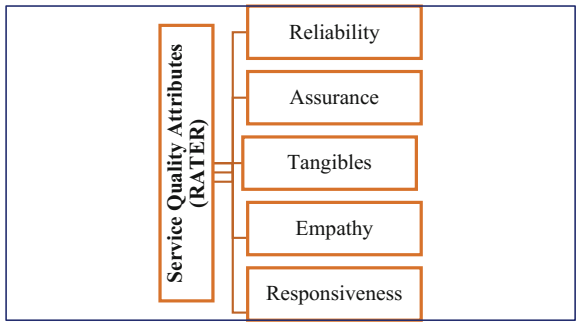

**Fig. 2** Service quality model

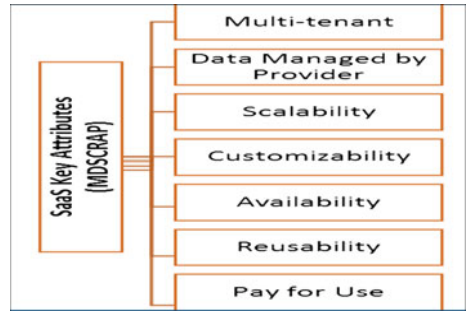

Fig. 3 Key features of SaaS

- *Data Maintained by Service Provider*: SaaS model of software arrangement provides service suppliers authorization solicitations anywhere to service users on demand. So that service suppliers will be additionally responsible for installation and data management. Hence, maximum data of clients are stored in the service provider's data center and maintained by them.
- *Scalability*: It is a desirable feature of the cloud services. Scalability means ability to handle increasing quantities of tasks or workloads. Service users can not have the control over resources. Service suppliers are solely liable for expanding services as per customer requests.
- *Customizability*: It is a capability used for services to be altered through service users, so that service users can utilize services effectively.
- *Availability*: The service users are capable of using SaaS in the cloud computing environment from a Web browser through the Internet. The customers are not having proprietorship to use the SaaS. That means the software have to be installed and run on the service supplier's server. This feature is one of the most critical in the SaaS usage.
- *Reusability*: This defines a capability of reuse of software essentials for building various applications. The main principle of cloud computing is to use again and again several kinds of services available on the internet. In cloud computing reusability is an essential feature of SaaS.
- *Pay-per-use*: The expenses of Software-as-a-services are purely based on the usage of service and are not related to purchase of ownership [1].

(d) SaaS Cloud Quality Model

Proposed quality model of SaaS is based on the two important aspects of SaaS: software quality and service quality. SaaS quality model is also involved with the key feature of SaaS. In this model all key features of SaaS are identified and are mapped to software product quality model as well as service quality model. Further some metrics to measure quality of SaaS are derived. The work flow of the model is shown in the Fig. 4.

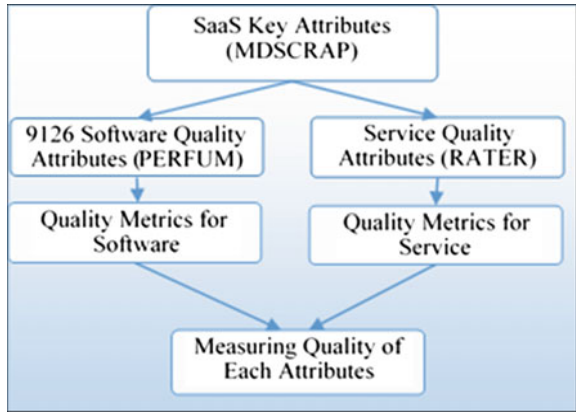

Fig. 4 Proposed quality model work flow

## 3 Results and Discussion

(1) Mapping Software Quality Attributes

The proposed quality model is used to map the SaaS key features with standard software product quality attributes as per ISO/IEC 9126. Mapping relationship between the key features of Software-as-a-service plus quality attributes of ISO 9126 standard are shown in the Fig. 5. The goal of this mapping is to empower the capacity of each key feature of SaaS and to have metric to measure its software product quality. Similarly the key features of SaaS are mapped with each quality attribute of software product quality model and derived related metrics as shown in the Fig. 7. This enables to measure the quality of each attribute of SaaS with respect to software.

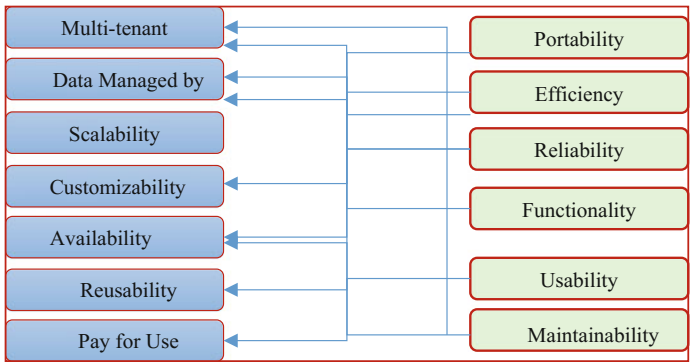

Fig. 5 Mapping software quality attributes

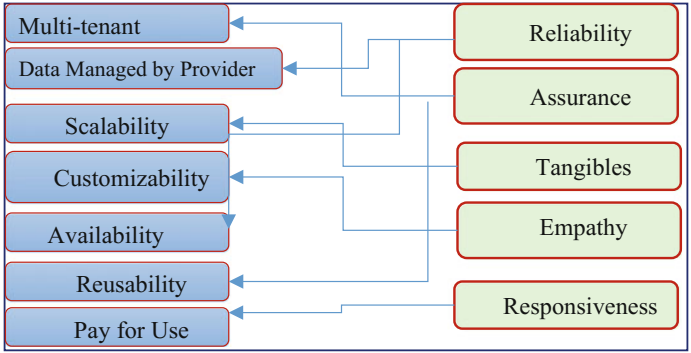

**Fig. 6** Mapping service quality attributes

(2) Mapping Service Quality Attributes

The SERVQUAL is a widely accepted evaluation model of service quality which is used as the basic framework [8] for service quality on the cloud computing environment so that all key features of SaaS are mapped with service quality attributes as shown in the Fig. 6.

(3) Derived Metrics

The goal of this mapping is to empower the capacity of each key feature of SaaS to have a metric to measure its service quality. Similarly for all key features of SaaS when mapped with each quality attribute of the service quality model and derived related metrics gives a good measure of quality. This facilitates the best possible way to measure the overall quality as well as quality of each attribute of SaaS with respect its service. Many metrics are derived to measure software and service quality of SaaS as shown in the Fig. 7.

- Maturity: Occurrence of failure of the software [9].
- Interoperability: Capability of software element towards interacting with further components [10].
- Suitability: Correctness to arrangement of purposes of software [9].
- Accurateness: Rightness of the functions [9].
- Recoverability: Capacity towards to revert back unsuccessful system to complete working system, including data plus network links [9].
- Understandability: Standardizes the effortlessness system functions can be understood and communicates to human being perceptual model to use easily [9].
- Changeability: Characterizes the amount of effort needed for code modification [9].
- Install ability: Illustrates struggle necessary to install software [9].

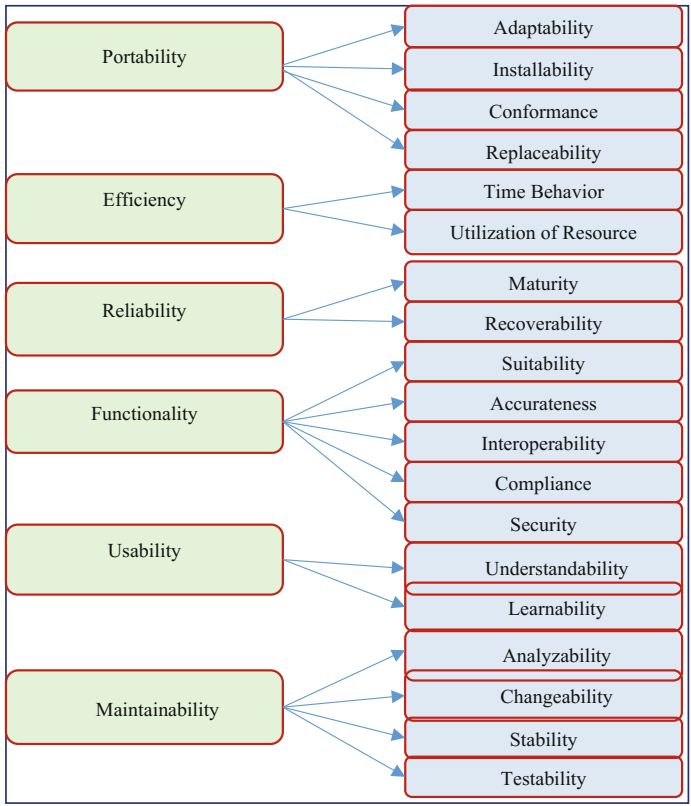

**Fig. 7** Derived metric for software quality attributes

## 4 Conclusion and Future Scope

The new proposed model has been introduced based on the key features of SaaS. All identified key features have been mapped with standard quality attributes of software product and service, as SaaS is the combination of both software products as well as service. With the help of mapping quality metric had been derived to evaluate a quality of SaaS. Further, it is also intended to implement automated tool by using this model to evaluate SaaS quality.

## References

1. J. W. J. Y. Lee, "A Quality Model for Evaluating Software-as-a-Service in Cloud Computing," in Software Engineering Research, Management and Applications, 2009. SERA '09. 7th ACIS International Conference on, 2009, pp. 261–266.
2. J. -M. Desharnais, "Analysis of ISO/IEC 9126 and 25010," 2009.

3. P. X. Wen and L. Dong, "Quality Model for Evaluating SaaS Service," in 2013 Fourth International Conference on Emerging Intelligent Data and Web Technologies, 2013, pp. 83–87.
4. Gajah, S. N. R. Sankranti, S. A. Wahab, N. J. A. Y. Norazira Abas, and S. N. A. M. Rodzi, "Adaptive of SERVQUAL Model in Measuring Customer Satisfaction towards Service Quality Provided by Bank Islam Malaysia Berhad (BIMB) in Malaysia," Int. J. Bus. Soc. Sci., vol. 4, no. 10, pp. 189–198, 2013.
5. A Parasuraman, V. A Zeithaml, and L. L. Berry, SERQUAL: A Multiple-Item scale for Measuring Consumer Perceptions of Service Quality, vol. 64. 1988, p. 28.
6. Dhanamma Jagli, Dr. Seema Purohit, Dr N. Subhash Chandra "SAASQUAL : A Quality Model for Evaluating SaaS on The Cloud," pp. 1–6, 2015.
7. Q. He, J. Han, Y. Yang, J. Grundy, and H. Jin, "QoS-driven service selection for mul-ti-tenant SaaS," Proc. - 2012 IEEE 5th Int. Conf. Cloud Comput. CLOUD 2012, pp. 566–573, 2012.
8. Z. Wang, N. Jiang, and P. Zhou, "Quality Model of Maintenance Service for Cloud Computing," 2015 IEEE 17th Int. Conf. High Perform. Comput. Commun. (HPCC), 2015 IEEE 7th Int. Symp. Cybersp. Saf. Secur. (CSS), 2015 IEEE 12th Int. Conf Em-bed. Software. pp. 1460–1465, 2015.
9. F. Beringer, "Software quality metrics and model," 2009.
10. IEEE Computer Society, "IEEE Standard for a Software Quality Metrics Methodology - IEEE Std 1061TM-1998 (R2009)," vol. 1998, 2009.